\documentclass[preprint,superscriptaddress,amsmath,amssymb]{revtex4}
\usepackage{graphicx}
\usepackage{dcolumn}
\usepackage{bm}
\usepackage{color}
\usepackage{natbib}

\begin{document}

\title{Dynamics of gravity driven three-dimensional thin films on hydrophilic-hydrophobic patterned substrates}

\author{\firstname{R.} \surname{Ledesma-Aguilar}}%
\email{rodrigo@ecm.ub.es}
\affiliation{Departament d'Estructura i Constituents de la
  Mat\`eria. Universitat de Barcelona,
Avinguda Diagonal 647, E-08028 Barcelona, Spain}
\author{\firstname{A.} \surname{Hern\'andez-Machado}}
\affiliation{Departament d'Estructura i Constituents de la
  Mat\`eria. Universitat de Barcelona,
Avinguda Diagonal 647, E-08028 Barcelona, Spain}
\author{\firstname{I.} \surname{Pagonabarraga}}
\affiliation{Departament de F\'isica Fonamental.  Universitat de
  Barcelona, Avinguda Diagonal 647, E-08028 Barcelona, Spain}
\date{\today}

\begin{abstract}
We investigate numerically the dynamics of unstable gravity driven three-dimensional thin liquid films on hydrophilic-hydrophobic patterned   
substrates.  
We explore longitudinally striped and checkerboard arrangements.  Simulations show that for longitudinal
stripes, the thin film can be guided preferentially on the hydrophilic stripes, while fingers develop on 
adjacent hydrophobic stripes if the width of the stripes is large enough.  On checkerboard patterns, 
the film develops as a finger on hydrophobic domains,  while it spreads laterally to cover the hydrophilic domains, providing a mechanism
to tune the growth rate of the film.    By means of kinematical arguments, we quantitatively predict the growth rate 
of the contact line on checkerboard arrangements, providing a first step towards potential techniques that 
control thin film growth in experimental setups.
\end{abstract}

\maketitle

\section{Introduction}

\begin{figure}
\centering
\includegraphics[width=0.35\textwidth]{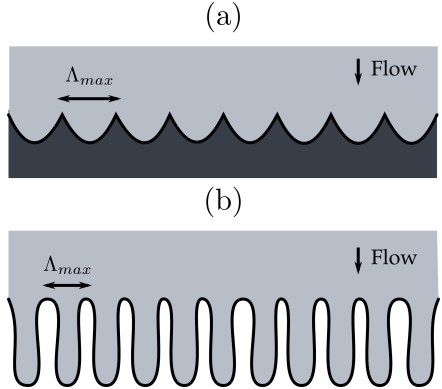}
\caption{Schematic top view of sawtooth and finger formation
in unstable driven liquid films on homogeneous substrates.  The contact line, 
shown in black, destabilizes and forms sawtooth 
structures on hydrophilic substrates (a) indicated by dark grey, 
while fingers grow on hydrophobic substrates (b), shown in white.  Each structure has a typical
periodicity, given by the most unstable wavelength of the front in the linear 
regime, $\Lambda_{max}$, which depends of the applied forcing, the inclination angle
of the substrate and the wetting properties of the fluid.\label{fig:Schematic}}
\end{figure}

Forced liquid thin films appear in many processes, such as solid coating, wetting of 
bio tissues and microfluidic flow orientation \cite{Bankoff01,Troian04,Marshall01}. For 
such driven films advancing on dry substrates, the forcing of the film triggers the 
destabilization of the contact line \cite{Brenner01,Silvi01,Jerret01,deBruyn01,Bankoff01}, which gives rise to 
interfacial structures of different shapes, as depicted in Fig.~\ref{fig:Schematic}, 
depending on the driving force, the wettability of the substrate, and, for gravity driven films, on the 
inclination angle of the substrate \cite{Bankoff01,Brenner01}.  
On hydrophilic substrates, the instability typically causes a deformation of the film that has the shape 
of saturated sawtooth structures, such as the ones shown in Fig.~\ref{fig:Schematic} (a), which do not 
grow in time, and that propagate along the forcing direction at the typical injection velocity $U$.  On the 
contrary, if the substrate is hydrophobic, the contact line breaks into steadily growing fingers, as in 
Fig.~\ref{fig:Schematic} (b), which propagate in the direction of forcing, growing at an intrinsic rate.  
Both sawtooth and finger structures have a typical transverse periodicity, which is characterized by the 
most unstable wavelength in the linear regime, $\Lambda_{max}$.   Again, this intrinsic lengthscale 
depends on the applied forcing, the inclination angle, and on the wetting properties of the substrate.  

Due to its relevance in technological and biological applications, \cite{Bankoff01,Troian04,Marshall01}, 
new insight on techniques for handling liquid films is needed. A promising scenario is that 
of exploiting substrate heterogeneity to manipulate the motion of thin films, specially in 
the emerging field of microfluidics \cite{Beebe01}, where a thin film geometry is a small friction 
alternative to common channels, as a medium to conduct microflows \cite{Dietrich01}.  Substrate
heterogeneity has proven useful for interface handling, as in the sorting of drops running on dry substrates 
\cite{Kusumaatmaja-Lang-2007}, or the creation of mixing domains in microchannels \cite{Balazs-Lang-2001}.

By profiting from heterogeneous substrates, our main motivation in this work will be to find ways in which thin films can be handled, orienting
them toward a preferred path and controlling the way in which contact line structures grow.  We shall perform a theoretical study, 
based on full three-dimensional hydrodynamical simulations, of the effect of imposing a hydrophilic-hydrophobic chemical pattern 
on the growth of the unstable contact line,  exploring a variety of pattern configurations to gain understanding on how to control 
the motion of the thin film, on top of the intrinsic time and lengthscales of the contact line instability. 

The problem of unstable contact lines in driven thin films has been actively studied during the 
last two decades.  Several experimental studies have addressed the problem 
of films driven by gravity on homogeneous substrates. In general, experiments are performed 
either by forcing a constant volume of liquid or by driving the film at constant injection rate.   
The constant volume configuration has been studied by Silvi and Dussan \cite{Silvi01}, Jerret and de Bruyn \cite{Jerret01}, and de Bruyn \cite{deBruyn01} using dry solids.  
They all agree in that sawtooth shaped structures are obtained on hydrophilic substrates, while 
fingers grow when using a hydrophobic substrate.  Veretennikov, Indeikina and Chuang \cite{Chuang02} observe the growth of fingers on a dry substrate, while for a prewet substrate sawtooth shapes are observed.  For films driven at a constant flux, Johnson, Schutler, Miksis and Bankoff \cite{johnson01} observe that sawtooth and finger 
structures can be obtained for the same fluid and substrate combination by varying the inclination angle.   

Theoretical efforts have focused on the early stages of contact line destabilization
and the subsequent nonlinear dynamics on homogeneous substrates.  Generally speaking, 
the framework is that of lubrication theory, in which the film dynamics are reduced 
to a two dimensional problem by profiting from the smallness of the film thickness, and which
is valid for hydrophilic substrates.  Wetting properties enter in the regularization of the spurious 
divergence of viscous dissipation at 
the contact line.  Standard approaches to this problem either relax the no slip 
boundary condition, thus allowing  for slip at the solid, or include a precursor film just 
in front of the macroscopic film.  Within the lubrication framework,
results for the linear stability of the contact line show that the instability is weakened 
by decreasing the inclination angle of the substrate \cite{Brenner02}, by increasing  the
 thickness of the precursor film \cite{Troian03} or by increasing the slip velocity \cite{Homsy01}.  
The main effect is contained in the three dimensional structure of the film at the proximity of the contact line, where the free surface is smoothed by surface tension and a capillary ridge is formed.  The thickness of the capillary ridge decreases 
as the wettability of the surface increases, or as the inclination angle of the substrate decreases.  As explained by Brenner \cite{Brenner01},  when a flat contact  line is perturbed, spatial variations of the
thickness of the ridge trigger transverse capillary flows from the troughs to the peaks of the perturbation.  As a result, the ridge thickens 
at the peaks  and the perturbation grows in time, the growth rate and band of unstable modes being determined by the 
thickness of the ridge.   At long times, nonlinear structures emerge.    Moyle, Chen and Homsy \cite{Homsy03} studied this regime numerically 
using the slip model,  while Kondic and Diez \cite{Diez01} used the precursor film model.    
These studies indicate that whatever the shape of the contact line, sawtooth or finger, there exists a 
natural length scale for finger or sawtooth spacing which is roughly determined by the most unstable 
wavelength predicted by linear theory.  

On perfectly homogeneous substrates, one expects that growing fingers are equally spaced 
because of the intrinsic length scale scale imposed by the most unstable wavelength.   
Experiments show that this ideal configuration is not easy to obtain,  
as finger spacing has a  high dispersion, of about 30\% \cite{Jerret01,deBruyn01}, due to, {\it e.g.}, substrate impurities.  
In an attempt to control finger spacing,  chemically patterned substrates have been studied both experimentally and numerically.     
Kataoka and Troian \cite{Troian04} studied the growth of fingers driven by thermocapillary 
stresses on alternated  low  and high energy longitudinal stripes.  Kondic and Diez \cite{Diez03,Diez04} performed  experiments 
and numerical simulations of the lubrication equations of gravity driven films on patterned substrates composed of low and high 
flow resistance longitudinal stripes.  Zhao and Marshall \cite{Marshall01} performed numerical simulations of the lubrication equations 
considering sinusoidal variations of fluid wetting properties. Both experiments and numerical simulations show that for such longitudinal 
patterns, the fluid advances preferentially over the domains that have a smaller flow resistance. As a consequence, the spacing between 
fingers can be controlled if the width of the stripes is comparable to the most unstable wavelength of the front.   For thinner stripes, 
surface  tension smooths the interface, creating wider structures. On the other hand, wider stripes eventually allow for the growth of  
unstable wavelengths, which generate additional fingers.    

In this paper we will study the orientation of the thin film on 
hydrophilic-hydrophobic patterns of sharply contrasted wetting properties, thus departing from the hydrophilic regime of previous 
numerical studies \cite{Diez03,Diez04,Marshall01}, analyzing the interplay between the typical lengthscale of the pattern and the intrinsic lengthscale of the instability. 
Apart from orienting the film on a prescribed path, another interesting problem is how to control the 
growth of the contact line structures triggered by the instability. Both experimental and theoretical 
results show that fingering can be suppressed by either decreasing the inclination angle of the substrate 
or by increasing its wettability sufficiently, as in these limits the contact line saturates to a sawtooth shape.   
Decreasing the inclination angle is effective only  for mostly wetting fluids; for non wetting fluids fingering is 
observed even at small inclination angles\cite{Silvi01,Jerret01,deBruyn01,Chuang02}.   Increasing surface 
wettability is then an appealing strategy to improve surface coverage.  A way to increase the wettability of 
the substrate is by imposing a chemical pattern on the solid.  Such an exploitation of substrate heterogeneity  
to control the growth of the film has not been studied previously.  This shall be a second question 
to be addressed in this work.

Overall, the instability poses two challenges when trying to handle thin films.  First, there is the presence of an
intrinsic lengthscale, $\Lambda_{max}$, which imposes the typical spacing between contact line structures.  
Secondly, there is an intrinsic timescale, which corresponds to the growth rate of the contact line.  
We shall therefore explore the effect of patterns where the typical time and lengthscales 
are comparable to $\Lambda_{max}$ and the typical growth rate of the film, respectively. 
Hence, in a hydrophilic domain, the front is expected to saturate to a sawtooth, while for 
hydrophobic domains contact line growth should occur.  

The rest of this paper is organized as follows. In section \ref{sec:goveqns} we present the hydrodynamical model and the lattice-Boltzmann integration algorithm that 
we use to perform the numerical simulations.  Section \ref{sec:Results} is devoted to the results of gravity driven 
thin films on different substrates.  In section \ref{sec:patterns} we study the formation of  sawtooth- and finger-shaped fronts on 
homogeneous substrates.  We follow by analyzing the growth of the contact line in different patterned substrates. Section \ref{sec:LS} is devoted
to longitudinal patterns of hydrophilic-hydrophobic stripes. In section \ref{sec:CB}  we consider arrangements of hydrophilic-hydrophobic domains 
in checkerboard patterns.  Finally, in section \ref{sec:DC} we discuss our results and present the conclusions of this work. 

\section{Diffuse interface model for driven thin films}
\label{sec:goveqns}
In this section we present the diffuse interface framework on which we 
rely to study thin film dynamics on chemically heterogeneous substrates. 
In a recent work \cite{Ledesma03}, we have studied the dynamics of thin films on 
homogeneous substrates in the aforementioned framework,  which
considers the hydrodynamics of two immiscible liquid phases coupled by a diffuse interface. 
Such a model allows for contact line dynamics and has been validated  against molecular 
dynamics simulations \cite{Qian01}.   Contact line dynamics 
emerges as a consequence of a diffusive mechanism at the contact line,  making the use of explicit 
boundary conditions at the contact line unnecessary  
\cite{Yeomans01,Ledesma01}.  This model has been used to study the dynamics of the contact line
between parallel plates on hydrophilic-hydrophobic domains by Wang {\it et al.} \cite{Qian02}.  Lattice-Boltzmann simulations, 
the powerful computational fluid dynamics algorithm that we shall use in this work, have been used to apply this model on a variety 
of situations. Briant and Yeomans \cite{Yeomans01} studied the motion of contact lines between shearing parallel plates.   The development of thin films in Hele-Shaw 
cells was studied by Ledesma-Aguilar {\it et al.}\cite{Ledesma01}. Three-dimensional effects on the Saffman-Taylor finger was studied by Ledesma-Aguilar {\it et al.} \cite{Ledesma02}.   
In Ref. \cite{Ledesma03} we have performed lattice-Boltzmann simulations of thin films flowing on homogeneous substrates
of arbitrary wetting properties.  Our results for the linear stability of the front show that the contact line is unstable to a wider band 
of perturbation modes as the equilibrium contact angle increases.  To validate our model, we  have shown that the previous lubrication 
theory results are approached for sufficiently small driving velocities. In addition,  quantitative agreement has been obtained between the 
lattice-Boltzmann thin film profiles and  those obtained by Spaid and Homsy \cite{Homsy01} for the slip-velocity lubrication model in the same 
small-velocity regime.   Long time dynamics show that a sawtooth saturates when the fluid wets the substrate, while fingers grow when 
forcing non-wetting fluids. In agreement with previous results \cite{johnson01,Diez01}, we have observed that sawtooth shaped 
fronts are obtained by decreasing the inclination angle of the substrate.

\subsection{Governing equations}
\begin{figure}

\begin{centering}

\includegraphics[width=0.45\textwidth]{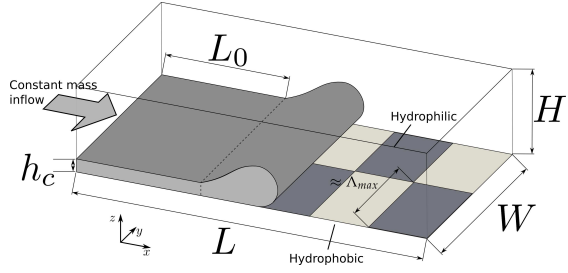}

\end{centering}

\caption{Schematic representation of the system geometry.  A thin film of thickness, $h_c$, width $W$, 
and initial length, $L_0$, is forced by gravity in the $x$ direction on an heterogeneous substrate composed of hydrophilic ($\theta_e=0^\circ$)
and hydrophobic ($\theta_e=90^\circ$) domains.  The film is surrounded by a fluid of much smaller viscosity.  The total system occupies a box of size 
$L\times W\times H$.  \label{fig:sysdiagram} }
\end{figure}

In this section we present the diffuse interface model that governs the dynamics 
of a gravity driven thin film in contact with a solid substrate. Figure \ref{fig:sysdiagram} 
shows a schematic representation of the system, which is composed of two immiscible 
liquid phases of different viscosities in contact with a solid wall. One phase corresponds to the driven 
thin film, while the other plays the role of a surrounding fluid of much smaller
viscosity.  Instead of considering that these phases are separated by a sharp interface,  we 
introduce a concentration variable, or order parameter, $\phi(\mathbf r)$, which has constant values in the bulk of each phase, 
$\phi = 1 $ for the thin film, and $\phi = -1$ for the surrounding fluid.  The concentration varies smoothly between these values across the interface.  
The dynamics of both phases follows by imposing that mass, momentum, and concentration are conserved in time.  

For an incompressible fluid, mass conservation gives the continuity equation
\begin{equation}
\boldsymbol \nabla \cdot \mathbf v = 0,
\label{eq:cont}
\end{equation}
where $\mathbf v(\mathbf r)$ is the velocity of the fluid.   Momentum conservation leads to the Navier-Stokes equations
\begin{equation}
\rho\left(\frac{\partial \mathbf v}{\partial t} + \left(\mathbf v \cdot \boldsymbol \nabla \right) \mathbf v \right) = -\boldsymbol \nabla P  + \left(\boldsymbol \nabla \cdot \eta \boldsymbol \nabla \right) \mathbf v +\rho \mathbf g -\phi \boldsymbol{\nabla} \mu. 
\label{eq:NS}
\end{equation}
The left hand side of equation (\ref{eq:NS}) contains the usual acceleration and inertial terms of the velocity
 field. The right hand side is composed of the usual pressure gradient term, where $P$ is the pressure
 field, the body force term, where $\rho$ is the fluid density and $\mathbf g$ is the acceleration due to gravity,  
and the viscous friction term of a Newtonian fluid, where $\eta(\phi)$ is the viscosity of the fluid, which depends
on the order parameter. The additional term in equation (\ref{eq:NS}) is the chemical force per unit volume of fluid, 
associated with the gradient of the chemical  potential $\mu(\phi).$   This force is important only where 
the concentration profile varies significantly, {\it i.e.}, at the diffuse interface.  In the sharp interface limit this interfacial forcing gives rise to the Young-Laplace 
condition (see Ref. \cite{bray} for details).

The dynamics of the order parameter is given by the conservation equation $\partial_t \phi + \mathbf v \cdot \boldsymbol \nabla \phi = -\boldsymbol \nabla \cdot \mathbf j,$ where the current,  $\mathbf j$, is given by $\mathbf j = -M \boldsymbol \nabla \mu.$  Hence, $\phi$ obeys the  convection diffusion equation 
\begin{equation}
\frac{\partial \phi}{\partial t} + \mathbf v \cdot \boldsymbol \nabla \phi = M \nabla^2 \mu,
\label{eq:CD}
\end{equation} 
where $M$ is a mobility.     

In equilibrium, the state of the system is determined by the free energy functional
\begin{equation}
F[\phi,\rho]=\int \mathrm d \mathbf{r} \left(V(\phi,\rho)+\frac{\kappa}{2}(\boldsymbol{\nabla} \phi)^2\right).
\label{eq:freeenergy}
\end{equation}
The first term in equation (\ref{eq:freeenergy}) is a volume contribution, $V(\phi,\rho)=A\phi^2/2+B\phi^4/4+\rho/3\ln\rho$, 
which allows for two phase coexistence via the $\phi$ dependent terms, and contains an ideal 
gas contribution given by the $\rho$ dependent term.  The second contribution to the free energy is the squared gradient term, 
which accounts for the
energy cost of spatial variations of the order parameter by a factor $\kappa$.   From this energy functional one can work out 
expressions for the chemical potential, $\mu$, the total pressure tensor $\boldsymbol{\mathcal P}$, the surface tension of the fluid-fluid 
interface, $\gamma$, the interface width, $\xi$,
and the equilibrium values of the order parameter in the bulk of the phases $\phi^* = \pm \phi_{eq}$. 
The chemical potential is given by,
$$\mu \equiv \frac{\delta F }{\delta \phi} = A\phi +B \phi^3 -\kappa \nabla^2\phi.$$
In this expression, $\mu$ corresponds to the energy cost of varying the concentration $\phi$, hence, it should be 
understood as the exchange chemical potential of the system.   The total pressure tensor, $\boldsymbol {\mathcal P}$,  
is composed of a density dependent contribution arising from the dependence of $V$ on $\rho$, and a ``chemical'' contribution, which depends 
on the order parameter,
$$
	\begin{array}{ccl}
\boldsymbol{\mathcal P} & = &P(\rho)\boldsymbol \delta + \left(\frac{1}{2}A\phi^2+\frac{3}{4}B\phi^ 4 -\kappa\left(\phi \nabla ^2\phi +\frac{1}{2}|\boldsymbol \nabla \phi|^2 \right)\right)\boldsymbol{\delta}\cr
&&  +\kappa\boldsymbol\nabla\phi\boldsymbol\nabla\phi,
\end{array}
$$
 where $P= \frac{1}{3}\rho$ and $\boldsymbol \delta$ is the diagonal matrix.  Thus, by including the $\rho$ dependent term
in equation (\ref{eq:freeenergy}) one recovers the behavior of an ideal gas in the case of a single homogeneous fluid 
in equilibrium, characterized by vanishing order parameter gradients.   Taking the divergence of the pressure tensor gives the 
force density acting on the fluid
$$-\boldsymbol \nabla \cdot \boldsymbol {\mathcal P } = -\boldsymbol \nabla P -\phi \boldsymbol \nabla \mu,$$
which clarifies the origin of the chemical term in equation~(\ref{eq:NS}). In this paper we will consistently use $B=-A$, from which  it follows that 
the  remaining equilibrium properties of the system are given by $\gamma=(-8\kappa A/9)^{\frac{1}{2}}$, $\xi= (-\kappa/2A)^{\frac{1}{2}}$ 
and $\phi^* = \pm 1$. 

\subsection{System geometry and boundary conditions}
We choose a system geometry to mimic the constant flux configuration of gravity driven thin films, as shown in 
figure \ref{fig:sysdiagram}.  We choose a rectangular domain of linear dimensions
$L \times W \times H$. Initially, the thin film occupies the volume $V_0=L_0\times W\times h_c$.    
The thin film is forced in the $x$ direction while the surrounding fluid is left to evolve passively.  
This is achieved by setting the gravitational force as $\rho \mathbf g  = \rho g_x \frac{1}{2}(\phi +1)\mathbf{\hat x}$.
Similarly, the viscosity is varied using the mixing rule $\eta(\phi) = \langle \eta \rangle( 1+ \delta \eta \phi)$, where 
$\langle \eta \rangle$ is the mean viscosity and $\delta \eta$ is the viscosity contrast.   
The viscosity of the thin film is then $\eta_{+1}=\langle \eta \rangle (1+\delta \eta)$, while the viscosity of the 
surrounding fluid is $\eta_{-1}=\langle \eta \rangle (1-\delta \eta)$. 

To complete the model, boundary conditions for $\mathbf v$ and $\phi$ have to be provided.   We choose the system geometry so that
the solid substrate is parallel to the $x-y$ plane and located at $z=0$ as shown in figure~\ref{fig:sysdiagram}.  The impenetrability of the solid is ensured by fixing $ v_z (x,y,z=0) = 0$, while stick boundary conditions are imposed at the wall, 
$ v_x (x,y,z=0) = v_y (x,y,z=0)=0.$   We fix periodic boundary conditions in the $y$ direction, {\it i.e.}, $\mathbf v(x,y=W,z)=\mathbf v(x,y=0,z)$, $\phi(x,y=W,z)=\phi(x,y=0,z)$. In the $x$ direction a constant mass inflow is kept by fixing $\partial_x \mathbf v (x=\{0,L\},y,z) =\mathbf 0$ and $\partial_x \phi (x=\{0,L\},y,z)=0.$  At the upper boundary, a shear free boundary condition is imposed by fixing $\partial_z \mathbf v (x,y,z=H) =\mathbf 0$, while a vanishing concentration gradient normal to the boundary is fixed by imposing $\partial_z \phi (x,y,z=H)=0.$ 

In equilibrium, a drop of fluid sits on the solid describing the shape of a spherical cap 
that intersects the solid boundary with an equilibrium contact angle $\theta_e$(for vanishing $\theta_e$ the spherical
cap is replaced by a film that covers the whole substrate).   The equilibrium contact angle is determined by Young's Law $-\gamma_{SL_{+1}}-\gamma\cos\theta_e+\gamma_{SL_{-1}}= 0$, 
where $\gamma_{SL_i}$ is the surface tension between the solid  and fluid $i$.  Hence, to include wetting effects in the model, 
it is necessary to account for the surface energies of the  solid-fluid boundaries. Within the diffuse interface formulation, we  
use a Cahn surface free energy $F_S[\phi_S]=\int_S f_S(\phi_S(\mathbf r))\mathrm d S,$ which is the 
integral along the solid-fluid surface, $S$, of the free energy per unit area, $f_S=C\phi_S$, that depends on the local 
value of the order parameter at the boundary, $\phi_S$.  In this expression, the parameter $C$ can be varied to 
obtain a prescribed equilibrium contact angle.   Minimizing the overall free energy gives the boundary
condition at the solid
\begin{equation}
\frac{\partial \phi}{\partial z}(x,y,z=0)=\frac{1}{\kappa}\frac{\mathrm d f_S}{\mathrm d \phi_S},
\label{eq:BC1}
\end{equation}
from which the equilibrium
contact angle is related to the model parameters by the expression \cite{Pagonabarraga02} 
$$\cos \theta_e = \frac{1}{2}\left[-\left(1-C(-\kappa A)^{-\frac{1}{2}}\right)^{\frac{3}{2}}+\left(1+C(-\kappa A)^{-\frac{1}{2}}\right)^{\frac{3}{2}}\right].$$  
The desired chemical pattern is obtained by imposing equation~(\ref{eq:BC1}) to fix the gradient
of the order parameter at the wall.

The diffuse interface model naturally allows for local slip at the contact line region.  As noted by Briant and Yeomans \cite{Yeomans01} and Qian {\it et al.} \cite{Qian01}, even though
a stick boundary condition is imposed to the velocity field, the diffusive term in equation~(\ref{eq:CD}) allows for motion
of the contact line.  Such diffusive flux has been studied, {\it e.g.}, by Denniston and Robbins \cite{Robbins01} in miscible displacements of binary fluids
using molecular dynamics.  In our model, the diffusion mechanism induces relaxation of the stick boundary condition in a length 
scale $l_{d}$.  Close to the contact line, at scales comparable to $l_d$, one observes slip, while for scales much larger than this lengthscale stick is recovered.  
Detailed studies of this length scale have been performed by Briant and Yeomans \cite{Yeomans01} and  Qian {\it et al.} \cite{Qian01}.

\subsection{Dimensionless numbers and units}
In order to relate our simulation units with experimental ones, 
let us briefly review the limit of the model in the lubrication regime.   

As explained before, in the sharp interface limit the usual continuity and Navier-Stokes equations are recovered from equations~(\ref{eq:cont})
and (\ref{eq:NS}), while the chemical potential term in  equation (\ref{eq:NS}) gives the Young-Laplace condition at the 
interface, 
$$\Delta P = -\gamma \mathcal K,$$
where $\Delta P$ is the pressure jump at the interface and $\mathcal K$ is the interface curvature.  

Due to the smallness of the film thickness, $h_c$, in the lubrication limit only the in-plane components of the velocity 
field, $v_x$ and $v_y$, are considered, and velocity gradients are assumed to occur primarily in the perpendicular 
direction, $z$.  Additional assumptions are that the flow takes place at small velocities and that it is stationary. Hence, the 
left hand side of equation~(\ref{eq:NS}) can be neglected.  Following these assumptions, it is possible to average the flow field in the $z$ 
direction to give 
\begin{equation}
\langle \mathbf v \rangle =-\frac{h^2}{3\eta_{+1}}(\gamma \boldsymbol \nabla \nabla^2 h - \rho \mathbf g),
\label{eq:lub}
\end{equation}
where $\langle \mathbf v  \rangle(x,y)$ is the two dimensional average velocity field, and $h(x,y)$ is
the local thickness of the film.   In this equation, the major contribution to the local pressure is 
capillary,  $P = -\gamma \mathcal K \simeq -\gamma \nabla^2 h.$

In this limit, it follows that the coordinates, velocity, and time in equation~(\ref{eq:lub}) can be rescaled as 
$$x^* =  \frac{x}{x_c},\qquad y^*= \frac{y}{x_c} ,\qquad (z^*,h^*)= \left(\frac{z}{h_c},\frac{h}{h_c}\right),$$
\begin{equation}
\langle \mathbf v  \rangle^* = \frac{\langle \mathbf v \rangle }{U},\qquad\mathrm{and}\qquad \frac{t}{t_c},
\label{eq:units}
\end{equation}
using units $x_c=h_c(3Ca)^{-\frac{1}{3}}$,  $U=h_c^2\rho g_x/(3\eta_{+1})$ and $t_c=x_c U^{-1}.$  In these expressions, the capillary number, $Ca$, measures the ratio between viscous and capillary forces, and is defined as $Ca=\eta_{+1} U /\gamma$.

\subsection{Lattice-Boltzmann algorithm}
To integrate equations~(\ref{eq:cont}), (\ref{eq:NS}) and (\ref{eq:CD}) numerically, we use LUDWIG, 
a lattice-Boltzmann parallel implementation for binary fluids \cite{Pagonabarraga02}.  Space is represented 
by a lattice of nodes that are connected by links.  The set of links determines a set of velocity vectors $\{\mathbf c_i\}$. 
Here we choose the cubic lattice D3Q19 model for the velocity set $\{\mathbf c_i\}$, which consists of nineteen velocity 
vectors (eighteen pointing towards nearest, next nearest, and next to next nearest neighbors) and one accounting for 
rest particles.  The lattice-Boltzmann method consists of the  integration of two linearized discrete Boltzmann equations
\begin{equation}
f_i(\mathbf r + \mathbf c_i ,t+1 ) - f_i(\mathbf r,t) = \frac{1}{\tau_f}\left(f_i(\mathbf r,t) -f_i^{eq}(\mathbf r,t)+F^f_i\right),
\label{eq:lbf}
\end{equation}
and 
\begin{equation}
g_i(\mathbf r + \mathbf c_i ,t+1 ) - g_i(\mathbf r,t) = \frac{1}{\tau_g}\left(g_i(\mathbf r,t) -g_i^{eq}(\mathbf r,t)\right),
\label{eq:lbg}
\end{equation}
where $f_i$ and $g_i$ are velocity distribution functions of the set of allowed velocities $\{\mathbf c_i\}$.     In these expressions, lattice units for 
length, time, and mass are fixed to $\Delta x = 1$, $\Delta t =1$, and $\Delta m =1$, respectively.

The dynamics of the distribution functions consists of two steps. First, the distribution functions undergo a collision step, which corresponds to the right hand side of equations (\ref{eq:lbf}) and (\ref{eq:lbg}), where they are relaxed to equilibrium distribution functions, denoted by $f_i^{eq}$ and $g_i^{eq}$.  The relaxation timescale of this process is given by the parameters $\tau_f$ and $\tau_g$.  We fix $\tau_g=1$, while  
$\tau_f$ is related to the fluid viscosity through $\tau_f=(6\eta+1)/2$.  The extra term $F^f_i$ corresponds to the external gravitational field \cite{ladd}.   After the collision step, the $f_i$ and $g_i$ are propagated to the neighboring nodes of the lattice according to the left hand side of equations~(\ref{eq:lbf}) and (\ref{eq:lbg}).

The mapping between the lattice-Boltzmann algorithm and the diffuse interface model presented above
follows from the definition of the moments of the distribution functions.  The $f_i$ are related to the fluid density and momentum by $\rho = \sum_i f_i$ and $\rho \mathbf v = \sum_i f_i \mathbf c_i,$ while the $g_i$ are related to the order parameter by $\phi = \sum_i g_i.$  The relaxation process ensure the conservation laws $\sum_i f_i = \sum_i f_i^{eq}$,  $\sum_i f_i \mathbf c_i = \sum_i f_i^{eq} \mathbf c_i$, and $\sum_i g_i = \sum_i g_i^{eq}$.   In the simulation runs we ensure that the fluid velocities are always smaller  than the speed of sound, $c_s = 1/\sqrt 3$; hence the flows can be considered  incompressible.
To complete the model one needs expressions for the equilibrium distribution functions and for the forcing term in equations~(\ref{eq:lbf}) and~(\ref{eq:lbg}).  
These expressions are obtained by expanding $f_i^{eq}$ and $g_i^{eq}$, as well as $F^f_i$, in powers of the velocity, and then by relating their higher order moments to the equilibrium properties of the diffuse interface
 model \cite{Pagonabarraga02}.   For our implementation the $f_i^{eq}$, $g_i^{eq}$ and $F^f_i$ read,
$$f_i^{eq} = \rho \omega_\nu\left(A_\nu^f + 3\mathbf v\cdot \mathbf c_{i} + \frac{9}{2}\mathbf v\mathbf v : \mathbf c_{i}\mathbf c_{i} -\frac{3}{2}v^2 + {\boldsymbol{\mathcal G}}^f:\mathbf c_{i} \mathbf c_{i}\right),$$
$$g_i^{eq} = \rho \omega_\nu\left(A_\nu^g + 3\mathbf v\cdot \mathbf c_{i} + \frac{9}{2}\mathbf v\mathbf v : \mathbf c_{i}\mathbf c_{i} -\frac{3}{2}v^2 + {\boldsymbol{\mathcal G}}^g:\mathbf c_{i} \mathbf c_{i}\right)$$
and
$$F^f_i = 4\omega_\nu\left(1-\frac{1}{2\tau_f}\right)\left(\mathbf f \cdot \mathbf c_i(1+\mathbf v \cdot \mathbf c_i)-\mathbf v \cdot \mathbf f\right).$$
Here, $\nu$ stands for the three possible magnitudes of the $\mathbf c_i$ set.  Coefficient values are
$\omega_0 = 2/9,$ $\omega_1=1/9$ and $\omega_{\sqrt{3}}=1/72;$ $A^f_0=9/2-7/2\mathrm{Tr}\boldsymbol{\mathcal P} ,$
$A^f_1 = A^f_{\sqrt{3}} = 1/\rho \mathrm{Tr}\boldsymbol{\mathcal P}$ and $\boldsymbol{\mathcal G}^f = 9/(2\rho)\boldsymbol{\mathcal P}-
3\boldsymbol{ \delta}\mathrm{Tr}\boldsymbol{\mathcal P};$  $A_0^g= 9/2-21/2\hat M\mu,$  $A_1^g = A^g_{\sqrt{3}} = 3\hat M\mu/\rho$ and
$\boldsymbol{\mathcal G}^g = 9/(2\rho)\hat M\mu(\mathbf{1}-\boldsymbol{\delta}),$ where $\mathbf{ 1}$ is the unit matrix.
 
 It can be shown that equations~ (\ref{eq:lbf}) and (\ref{eq:lbg}) converge to the hydrodynamic
 equations~ (\ref{eq:cont}), (\ref{eq:NS}), and (\ref{eq:CD}) by performing a Chapman-Enskog expansion.  This procedure has 
been carried out elsewhere \cite{Yeomans02}.  

Boundary conditions in the lattice-Boltzmann scheme are implemented by fixing the distribution functions
at the boundaries. To implement stick and no flow conditions at the solid, bounce back rules 
are implemented at the solid nodes \cite{ladd}.  Periodic boundary conditions are obtained as usual, {\it e.g.}, 
for the $f_i$ at $y=W$,  $f_i(x,y=W,z)=f_i(x,y=0,z).$  Shear free and vanishing concentration gradient conditions are 
obtained in a similar way.  The Cahn wetting boundary condition given by equation~(\ref{eq:BC1}), is obtained by extrapolating the 
order parameter to the wall nodes in order to fix a prescribed gradient in the normal direction to the wall \cite{Pagonabarraga02}.

Throughout this paper, the diffuse interface model parameters are fixed to the following values (in lattice units): $\rho=1$, $\langle \eta \rangle = 0.1$,  $\delta \eta = 0.9$,  $\gamma=0.0023$,  $\xi=0.57$, and $M=16$.   These values generally ensure numerical stability in the simulations. 
The interface thickness, $\xi=0.57$, has been tested previously and proved not to give rise to important lattice artifacts \cite{pagonabarraga,Cates-PTRSAMES-2005}.  We will consider two equilibrium contact angles, $\theta_e=0^\circ$, and $\theta_e=90^\circ,$ which correspond to $C=0.00167$ and $C=0$ respectively.
A suitable value for the film thickness,  which is small enough to keep reasonable computational costs and large enough to 
ensure the relaxation to the volume values of the order parameter needed for proper hydrodynamic resolution, is  $h_c=8$.  The mean velocity of the film is fixed to $U=0.005$ by choosing 
the gravitational acceleration as $g_x=3\eta_{+1}U/(\rho h_c^2)$.   Using these values, the capillary number is $Ca=0.41$.

\section{Homogeneous Substrates} 
\label{sec:Results}

The intrinsic time and lengthscales in forced thin films 
are set by the film growth rate, $\dot L(\theta_e)$, and by the most 
unstable wavelength of the linked to the contact line 
instability, $\Lambda_{max}(\theta_e)$, which depend on the
wetting properties of the substrate, determined by the equilibrium contact angle, $\theta_e$.  
In order to explore the 
effect of substrate heterogeneity on the dynamics of the film, 
we must first explore the dynamics on homogeneous substrates, 
to characterize $\dot L $ and $\Lambda_{max}$. 
In this section we describe the dynamics of thin films on
two separate homogeneous substrates with equilibrium contact 
angles $\theta_e = 0 ^\circ $ and $\theta_e = 90^\circ$, 
respectively.  Our aim is to characterize the linear instability by measuring the 
dispersion relation on each substrate, and to characterize
the base structures that emerge at long times. 

\subsection{Linear Stability}
\label{sec:patterns}
\begin{figure}

\begin{centering}

\includegraphics[width=0.45\textwidth]{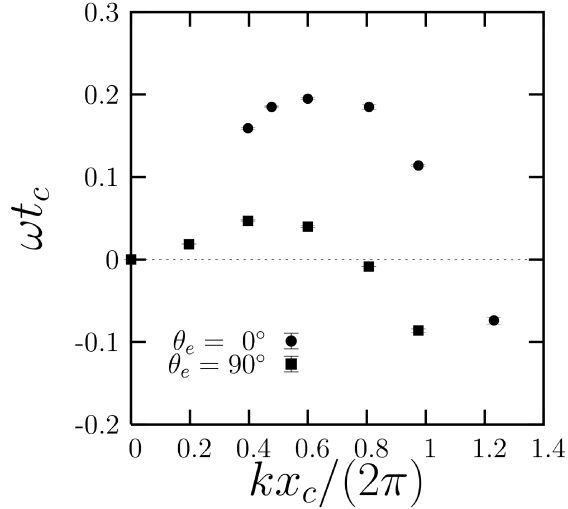}

\end{centering}
\caption{The growth rate, $\omega$,  of a transverse perturbation to the contact line as a function of the perturbation mode, $k$,  on    
homogeneous substrates with different wetting properties obtained from lattice-Boltzmann simulations.  
The growth rate has a maximum that depends on the equilibrium contact angle,  $\theta_e$.  The maximum growth rate is larger 
if the substrate is hydrophobic.   The corresponding most unstable wavelength, $\Lambda_{max}(\theta_e)=2\pi/k_{max}$, is smaller 
the more hydrophobic is the substrate.    The error bars correspond to the standard deviation of the fit performed to estimate $\omega$ from the
time evolution of the contact line amplitude.
  \label{fig:disp_rel}}
\end{figure}
\begin{figure*}

\includegraphics[width=0.05\textwidth]{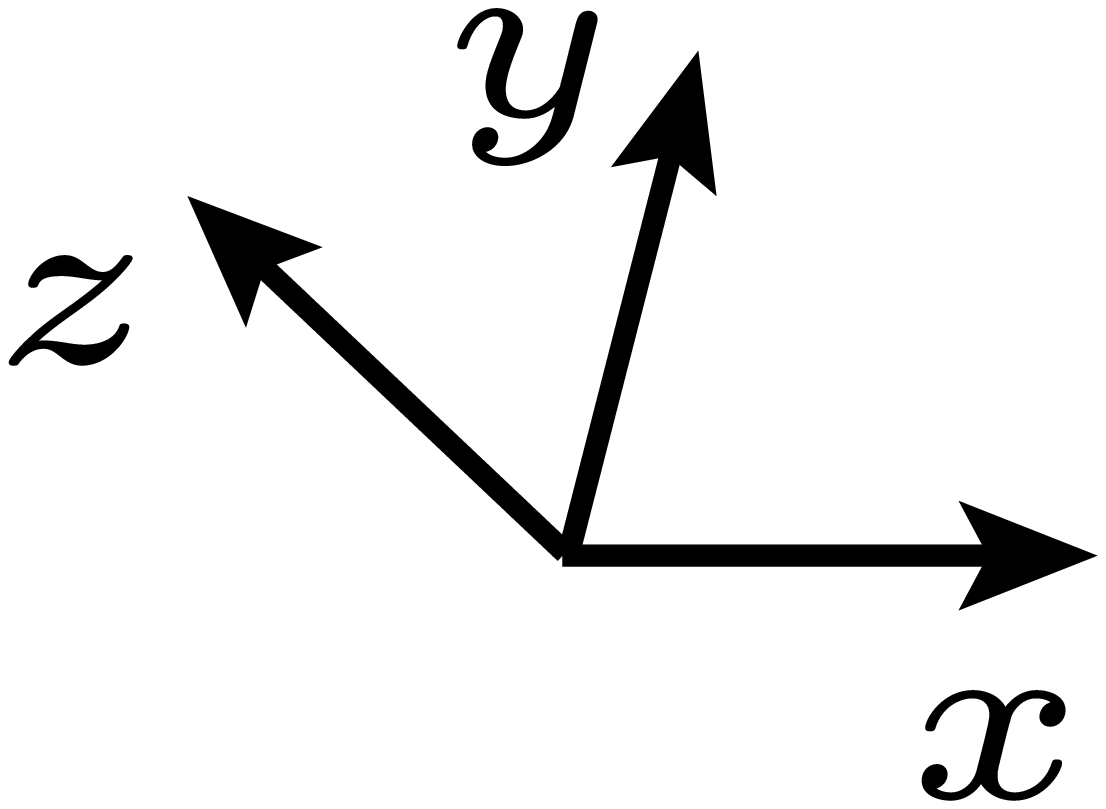}

\begin{centering}

\includegraphics[width=0.85\textwidth]{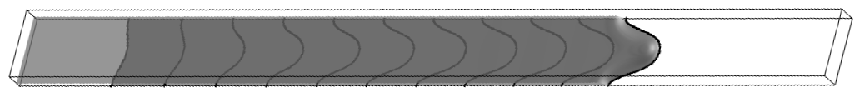}\\

(a)\\

\includegraphics[width=0.85\textwidth]{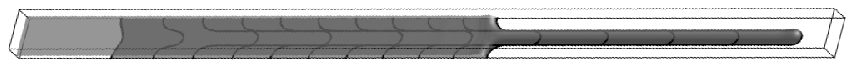}\\

(b)

\end{centering}

\caption{Evolution of unstable fronts in substrates of different wettabilities that originate from a slightly perturbed 
front at $t/t_c=0$. (a) Saturated sawtooth: $\theta_e=0^\circ$ and $W/x_c=13$. (b) Growing finger: $\theta_e=90^\circ$ and $W/x_c=8$.  The width of each system is chosen to match the most unstable wavelength as measured from figure~\ref{fig:disp_rel}, 
depending on the wetting properties of the substrate. The plot shows the contact line position at time intervals of $\Delta t/t_c = 13.4$, starting at $t/t_c=0$.  
For the last  time, $t/t_c=134$, the full three dimensional interface is shown. \label{fig:hsubs}}
\end{figure*}
 
A numerical linear stability analysis of the front can be performed by following the evolution of an initially flat contact line that is perturbed 
transversely with a single mode $k/k_c$, where $k_c=2\pi/x_c$.  At short times, the perturbation either grows or decays 
exponentially with a growth rate, $\omega(k)$.    We follow the evolution of single mode perturbations of small initial 
amplitude $A_0/W\approx 10^{-2}$, and focus on timescales where the amplitude shows an exponential dependence
on time to extract the growth rate of the perturbation.  The resulting dispersion relations for each front are 
shown in  figure~\ref{fig:disp_rel}.  The band of modes over which the front is unstable is much narrower for 
$\theta_e=0^\circ$ than for $\theta_e=90^\circ$.  This is due to  the thin film structure near the contact line, which 
tends to accumulate more mass when the fluid is in contact with hydrophobic substrates than with 
hydrophilic ones.  As a result, fronts on hydrophilic substrates are stable to longer wavelengths and 
have smaller growth rates.   In a recent work \cite{Ledesma03}, we have validated our numerical 
method by comparing the linear stability results of the diffuse interface model to those of lubrication theory.  
A direct comparison is only possible for small dynamic angles and small capillary number, $Ca$.  Given that 
our model allows for slip at the contact line, it is natural to compare
with the results obtained by Spaid and Homsy \cite{Homsy01}, for which the contact line is allowed to slip by imposing a Navier boundary condition
at the solid.  As explained in the aforementioned paper, our results approach 
the lubrication theory results for small $Ca$.

\subsection{Sawtooth and Finger formation}

 Figure~\ref{fig:disp_rel} shows  that the most unstable wavelengths are $\Lambda_{max}(\theta_e=0^\circ)/x_c\approx 14$ and $\Lambda_{max}(\theta_e=90^\circ)/x_c\approx 8$. These wavelengths set the typical length scale of the contact line shape at long times.  
Taking into account this periodicity, in the following we will reduce computational costs by considering system sizes whose width is of the order 
of $\Lambda_{max}(\theta_e)$.  Figure~\ref{fig:hsubs} shows the long time evolution of the most unstable wavelengths for $\theta_e=0^\circ$ and $\theta_e=90^\circ$.   We observe that after a transient, the thin film saturates to a sawtooth shape on the hydrophilic substrate, while on the hydrophobic substrate a finger grows at a steady rate. 
Given that the growth rate of the finger is constant, the substrate is not covered evenly, and a dry region is left to be covered by the trailing edge.  
This contrasts with  the dynamics in the hydrophilic substrate, on which the sawtooth saturates to a finite amplitude 
and propagates as a whole with a velocity dictated by the injection rate. 
Intuitively, one can understand the reason why a finger grows and a sawtooth saturates by examining the structure of the thin film.  
For unstable fronts,  the amplitude of a perturbation grows due to an uncompensated distribution of the thickness of the ridge.   
As the perturbation grows, the contact line becomes increasingly curved, thus increasing the strength of the restoring capillary
force.  If the thickness of the capillary ridge is small, as on hydrophilic substrates, the contact line 
can reach a sufficiently large curvature, and the distribution of the ridge thickness is balanced by surface tension.  Thus, 
the contact line saturates to a sawtooth shape.  In contrast, if the thickness 
of the ridge is large, as on hydrophobic substrates, the curvature of the contact line is always insufficient 
to balance the driving force, and fingers grow at a rate dictated by the distribution of the ridge thickness along the front.  We have demonstrated this effect
previously \cite{Ledesma03}, where we have shown that the growth rate of any sawtooth or finger can be collapsed to a 
universal curve, as a function of the difference of the thickness of the ridge between the leading and trailing edges of the contact line, 
independently of the capillary number, or the wetting properties or inclination angle of the substrate.  Static and dynamic fluid rivulets are known 
to be unstable to wave lengths which decrease with substrate hydrophobicity, gravity and  fluid flow~\cite{Davis-JFM-80,Dietrich-POF-06, Kondic-POF-09}. 
We have never observed the destabilization of a finger for the regimes we have explored, but we cannot rule it out; such a possibility requires a more systematic analysis.

\begin{figure}

\begin{centering}

\includegraphics[width=0.45\textwidth]{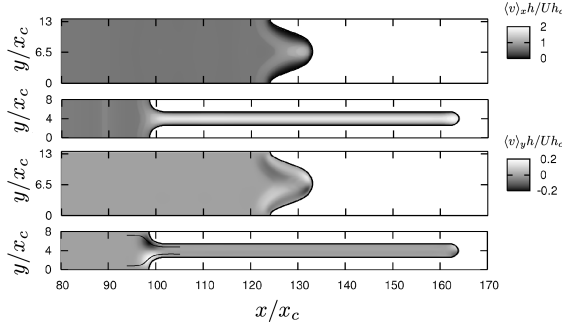}

\end{centering}

\caption{Grey scale maps of the average flow within the film for a steady sawtooth ($\theta_e=0^\circ$) and a finger ($\theta_e=90^\circ$) at $t/t_c=134$, starting from a slightly perturbed front at $t/t_c=0$.  The width of the system is fixed to $W/x_c=13$ for the hydrophilic substrate, and $W/x_c=8$ for the hydrophobic substrate. Top panels: the flow in the $x$ direction, $\langle v \rangle_xh/(Uh_c)$,  is homogeneous far from the contact line, and increases sharply on the body of the finger.  The sawtooth shows a small increase, due to the
variation of the film thickness. Bottom panels: the lateral flow, $\langle v \rangle_yh/(Uh_c)$, shows clearly that both the finger and the sawtooth 
are fed by transverse flows. For the finger, streamlines are computed close to the trailing edge and show that particles flow to the body of the finger as they approach this region.   
\label{fig:meanflows}}
\end{figure}

To characterize the flow on homogeneous substrates, we compute the local mean velocity by averaging the velocity field, $\mathbf v$, in the $z$ direction:
$\langle \mathbf v \rangle =1/h\int_0^h \mathrm{d}z \mathbf v$. Figure~\ref{fig:meanflows} shows 
plots of $\langle \mathbf v \rangle h$, which is a measure of the local mass flow, for both hydrophilic and hydrophobic substrates.  The top panels in
 figure~\ref{fig:meanflows} show the $x$ component of the flow. For the sawtooth, we observe only a slight increase of the flow near its tip. This is 
a signature of the balance between the driving force caused by the variation
of mass along the contact line, and the restoring surface tension. For a finger, this balance cannot be sustained, and
a sharp flow increase in the body of the finger is observed.  As shown in  the bottom panels in figure~\ref{fig:meanflows}, there exist two stationary regions that feed 
both the sawtooth and the finger.  For the sawtooth, these regions extend along the whole contact line structure 
sustaining its stationary shape.  For a  finger, they are located at the trailing edge of the contact line and fix a 
constant flow of mass that causes the finger to grow steadily (see, {\it e.g.}, the streamlines depicted at the lowermost panel 
of the figure).

\section{Heterogeneous Substrates}

In the previous section we presented results for sawtooth and 
finger growth on homogeneous substrates.  In both hydrophilic and
hydrophobic substrates, the lengthscale of instability is characterized 
by the most unstable wavelength of the linear regime, 
$\Lambda_{max}(\theta_e)$, which sets the typical spacing of both 
sawtooth and finger structures.  

In this section we explore the dynamics of the thin film on different 
hydrophilic-hydrophobic heterogeneous substrates.   Our aim is 
to examine the effect of a hydrophilic pattern imposed on a hydrophobic
substrate on the dynamics of the film.   Furthermore, taking into account 
the periodicity of the front, we will focus on patterns of characteristic lengthscales 
which are comparable to $\Lambda_{max}(\theta_e=90^\circ)$.   

\subsection{Longitudinal stripes}
\label{sec:LS}
\begin{figure}

\begin{centering}

\includegraphics[width=0.45\textwidth]{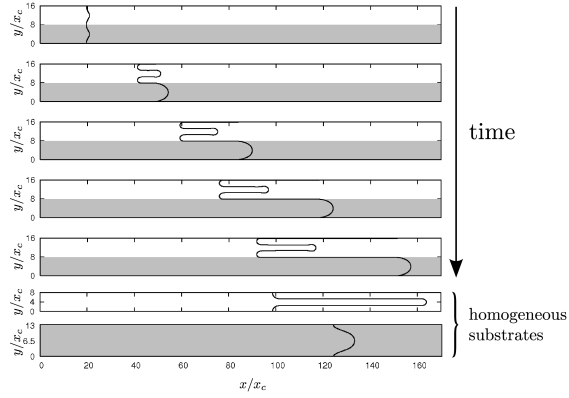}

\end{centering}
\caption{Contact line growth on a pattern composed of hydrophilic-hydrophobic stripes oriented in the 
direction of flow.  The dark and light stripes correspond to the hydrophilic (grey) and hydrophobic (white) substrates, respectively. 
Their width, $w$, equals the most unstable wavelength observed for the hydrophobic substrate, $\Lambda_{max}(\theta_e=90^\circ)/x_c\approx 8.$
The first five panels show the contact line at time intervals of $\Delta t/t_c=33.5$.  The first panel shows the perturbed contact line at $t/t_c=0$. The last three panels show the effect of the longitudinal pattern by comparing the contact line on the patterned substrate to the heterogeneous and homogeneous substrates with the same initial condition at $t/t_c=134$.  Wetting properties are fixed to $\theta_e=0^\circ$ at the grey zones, 
while at the white zones $\theta_e=90^\circ$.
\label{fig:test3}}
\end{figure}

In many situations, such as in microfluidic
flow guiding and filamenting \cite{Troian04,Dietrich01}, it is desirable to 
orient the growth of the contact line on a prescribed direction.
For substrates with spatially varying wetting properties, the flow is expected 
to orient toward the more hydrophilic regions due to their lower flow resistance.
Meanwhile, the front is expected to destabilize on the hydrophobic stripes if their 
width is of the order of the most unstable wavelength, $\Lambda_{max}(\theta_e=90^\circ).$ 
Hence, our main interest will be to examine the effect of the longitudinal pattern 
for different stripe widths.
In this section we analyze the growth of the contact line on longitudinal patterns of 
hydrophilic-hydrophobic properties.    

We first impose a pattern of alternating hydrophilic-hydrophobic stripes 
of equal width $w$, oriented in the longitudinal direction. Given that the periodicity 
of the front is determined by $\Lambda_{max}/x_c\approx 8$, we consider the effect 
of increasing the stripe width from $\Lambda_{max}$, choosing the widths 
$w/x_c=8$, $w/x_c=11$, and $w/x_c=13$.  Due to the periodicity of the front, larger 
widths corresponding to $w > 2\Lambda_{max}$ are not explored. 
As in the homogeneous case, we explore the evolution of only one mode, for which 
the corresponding wavelength is fixed as $\Lambda=w$.  

Figure~\ref{fig:test3} shows the evolution of the front for stripes of 
width $w/x_c=8$.  Given that for $\theta_e=90^\circ$, $\Lambda_{max}/x_c\approx 8$, 
the front destabilizes as it moves over the hydrophobic 
stripe and grows steadily.  In contrast, when in contact with the hydrophilic 
stripe, the front bends and saturates to a sawtooth shape.    The last three plots
in figure~\ref{fig:test3} show the contact line for the heterogeneous and 
homogeneous substrates at the same time.  By comparing these plots 
we notice that for the same injection rate the length of the contact line on the hydrophobic stripe is smaller than the length of the
finger that would have grown on a homogeneous substrate. 
In addition, its tip is located at a less advanced position.  In contrast, the sawtooth advances 
more rapidly on the heterogeneous substrate than on the homogeneous one.  

\begin{figure}

\begin{centering}

\includegraphics[width=0.45\textwidth]{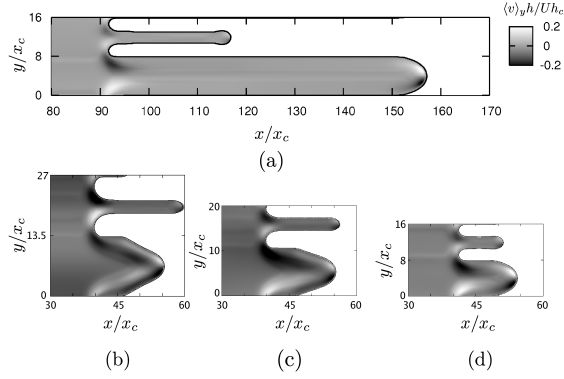}

\end{centering}

\caption{Grey scale maps of the $y$ component of the local mass flow  for a sawtooth and 
a neighboring finger in a pattern composed of longitudinal stripes.  (a) Steady configuration at $t/t_c=134$ for stripes of width $w/x_c=8$.  Early finger growth at $t/t_c=33.5$ for (b) $w/x_c=13$, (c) $w/x_c=11$ and (d) $w/x_c=8$ for an initial condition consisting of a slightly perturbed contact line with a transverse perturbation mode $k=2\pi/w$.   The effect of the striped pattern is to decrease the finger growth rate on the hydrophobic stripe as the width of the stripe decreases.  
\label{fig:meanflowslstripes}}
\end{figure}

To examine this partial inhibition of the growth on the hydrophobic stripe caused by the longitudinal 
pattern, we compute the $y$ component of the
flow,  which we show in figure~\ref{fig:meanflowslstripes}(a) for $w/x_c=8$.  The grey scale in 
this plot matches that of figure~\ref{fig:meanflows}, which corresponds
to a homogeneous substrate.  The flow field exhibits essentially the same regions of 
transverse flow as in the homogeneous case. These regions are stationary and only translate 
forward as time proceeds.  Therefore, the inflow to each stripe is constant in time. 
Nonetheless, by comparing the intensity of the regions between figure~\ref{fig:meanflows} and figure~\ref{fig:meanflowslstripes}(a)
one can appreciate that the inflow to the hydrophobic stripe is smaller for the heterogeneous substrate, while there is
a clear increase in the flow to the hydrophilic stripe, which makes the sawtooth propagate with 
a larger velocity compared to the homogeneous case.  

This effect is what causes the contact line on the hydrophobic stripe to have a smaller growth rate compared to the homogeneous situation. Nonetheless, 
given that transverse flows originate in the vicinity of the boundary between adjacent stripes,  
the effect should be observable only when the stripe width, $w$, is small enough. Figures~\ref{fig:meanflowslstripes}(b), \ref{fig:meanflowslstripes}(c), and 
\ref{fig:meanflowslstripes}(d) show flow maps in the $y$ direction corresponding to $w/x_c=8$, $w/x_c=11$ and $w/x_c=13$ 
at the early growth stage.   On the hydrophobic stripes, contact line growth is inhibited more effectively for narrow stripes, as 
can be seen from the intensity of the mass supplying regions depicted in the figure.  
For $w/x_c=13$, the effect is almost suppressed. The inflow to the hydrophobic stripe thus becomes 
almost equal to that of the homogeneous case. For much wider stripes, where the front is expected to 
break into several fingers, only those growing next to the hydrophilic stripes are expected to decrease 
their growth rate, while those located far from these boundaries should grow essentially as if they were 
in a homogeneous substrate. Anyhow, the film propagates faster on the hydrophilic 
stripes, so partial orientation of the film is achieved.   

\begin{figure}
\begin{centering}

\includegraphics[width=0.45\textwidth]{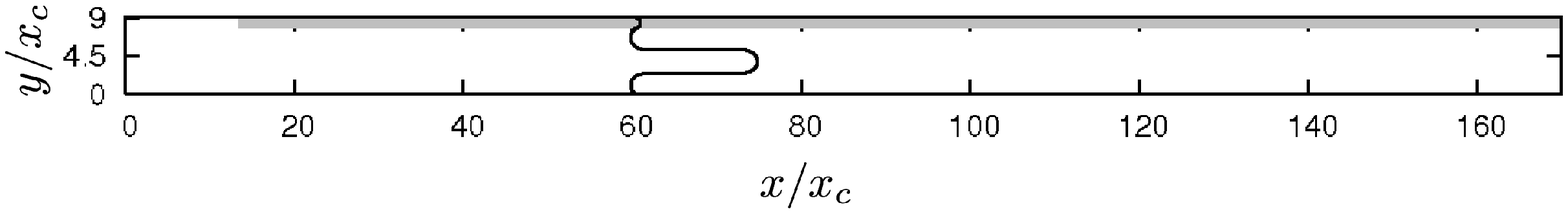}\\

(a)\\ 

\includegraphics[width=0.45\textwidth]{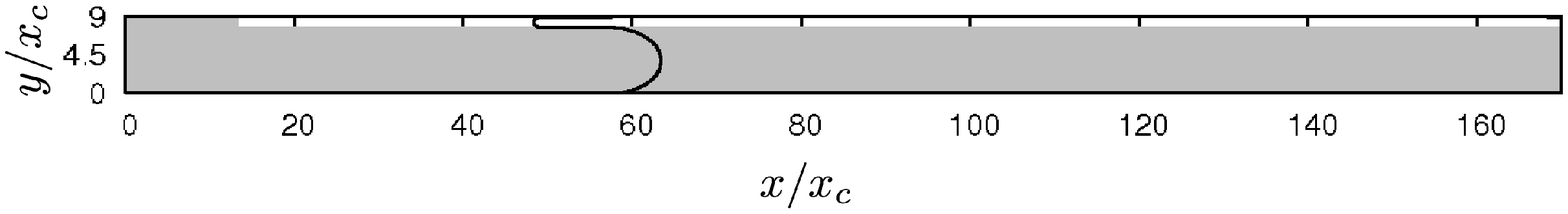}\\

(b)

\end{centering}

\caption{Contact line growth in wide/thin longitudinal stripe arrangements of hydrophilic (grey) or hydrophobic (white) properties. 
(a) For a wide hydrophobic stripe in contact with a thin hydrophilic stripe the contact line destabilizes on the hydrophobic domain, 
while the trailing edge advances on the hydrophilic domain. (b) On a wide hydrophilic stripe, the contact line forms a sawtooth,  
while the trailing edge of the contact line advances much more slowly on the thin hydrophobic domain. Wetting properties are fixed to $\theta_e=0^\circ$ at the grey zones, while at the white zones $\theta_e=90^\circ$. \label{fig:lstripes_small}}
\end{figure}

Having characterized the dynamics on stripes of  equal widths, it is straightforward to explore patterns 
where the stripes have disparate widths.   
In figures~\ref{fig:lstripes_small}(a) and \ref{fig:lstripes_small}(b) 
we show the evolution of the contact line on a very thin hydrophilic (resp. hydrophobic) stripe that is placed next to a wide hydrophobic (resp. hydrophilic)
stripe.  In both cases, the wide stripe has a width equal to $\Lambda_{max}$, 
while the thin stripe is small enough to avoid the growth of the contact line.   
Figure~\ref{fig:lstripes_small}(a)  shows that, in contrast with the case of wide neighboring stripes, in this limit the substrate structure perturbs weakly the contact line growth, by increasing the velocity of the trailing edge
 because of the  low resistance offered by the hydrophilic stripe.  In contrast, figure~\ref{fig:lstripes_small}(b) shows that the profile
has almost converged to the sawtooth expected on a homogeneous hydrophilic substrate, but the trailing edge in the thin hydrophobic 
stripe decouples from the trailing edge of the sawtooth, and advances with a much smaller velocity.   As a result, the film is oriented
selectively on the hydrophilic channel, given that the width of the hydrophobic stripe remains much smaller than $\Lambda_{max}$. 
This limit is precisely the one reported experimentally by Kataoka and Troian \cite{Troian04}, and by Kondic and Diez \cite{Diez03,Diez04} 
and Zhao and Marsall \cite{Marshall01}.   A particularly  interesting feature of this regime is that one can regard the 
growth of the contact line as a fingering process, which emerges as a consequence of the longitudinal pattern, and not because
of the linear instability of the contact line.  As such, it is possible to control the width of the emerging fingers by tuning the width 
of the hydrophilic stripes.  For large enough hydrophilic stripe widths, the contact line is expected to deform, giving rise to sawtooth 
structures.  However, this is expected to modified the morphology of the front only, and not to alter the fingering imposed by the longitudinal
pattern.  

\subsection{Checkerboard pattern}
\label{sec:CB}
\begin{figure}
\begin{centering}

\includegraphics[width=0.45\textwidth]{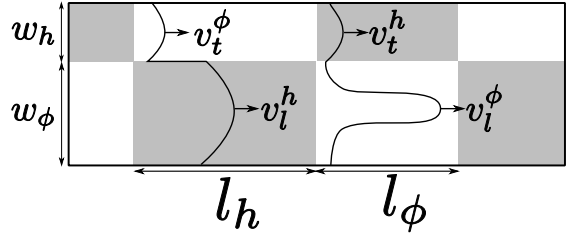}

\end{centering}
\caption{Geometry of the asymmetric checkerboard pattern.  The pattern is composed of alternated hydrophilic (grey)  and 
hydrophobic (white) domains.  In general, the width of the upper and lower domains is different, and the width of the lower domains 
is large enough to allow for the growth of unstable wavelengths that give rise to contact line growth on the hydrophobic domains and saturation 
to a sawtooth shape on the hydrophilic domains.  The lengths of hydrophilic and hydrophobic domains, $l_h$ and $l_\phi$,
can be chosen to fix a given fraction of hydrophilic covering.   Wetting properties are fixed to $\theta_e=0^\circ$ at the grey zones, 
while at the white zones $\theta_e=90^\circ$.  \label{fig:cbpattern}}
\end{figure}

\begin{figure}
\begin{centering}

\includegraphics[width=0.45\textwidth]{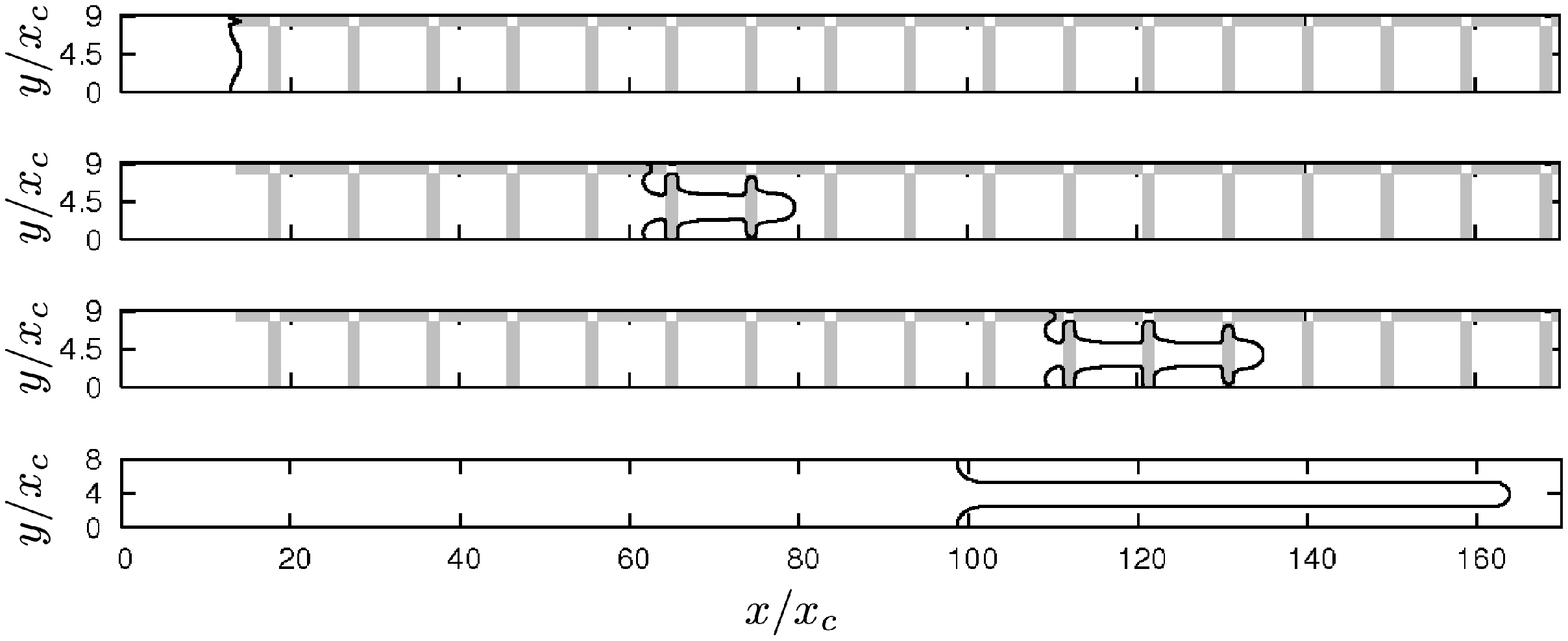}\\

(a)\\

\includegraphics[width=0.45\textwidth]{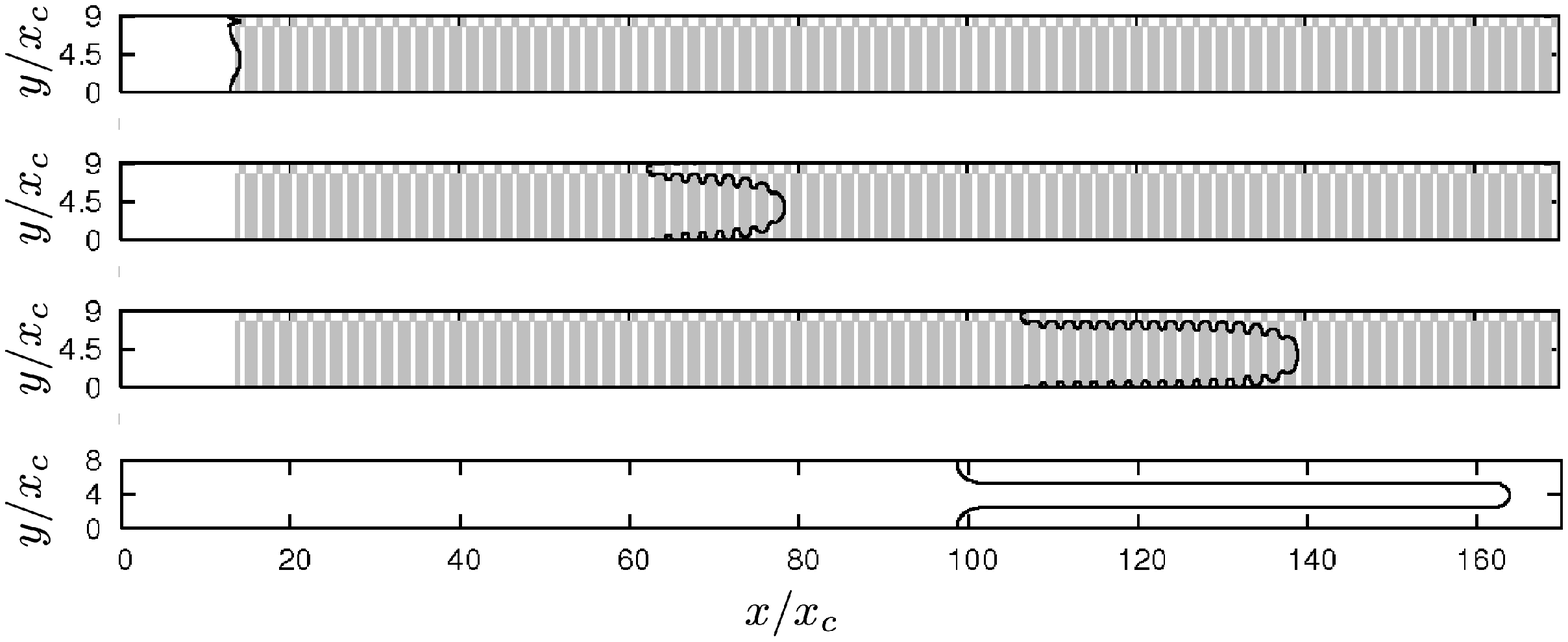}\\

(b)\\

\end{centering}
\caption{Contact line growth on asymmetric checkerboard patterns.  Patterns are fixed by choosing $l_h/x_c=1.3$ and (a) $f_{CB}=0.25$, and (b) $f_{CB}=0.62$.    
The first plot corresponds to a transverse perturbation of the contact line at $t/t_c=0$.   Subsequent plots are taken at time intervals of $\Delta t /t_c= 67.$  At the bottom of (a) and (b) we show the finger that grows in a homogeneous hydrophobic substrate for the same time as the last contact line snapshot of the checkerboard pattern.  By comparing (a) and (b),  it is observed that the length  of the growing contact line, $L$, is smaller if the fraction of hydrophilic domains is small. 
 \label{fig:cbpattern1}}
\end{figure}

\begin{figure}
\begin{centering}

\includegraphics[width=0.45\textwidth]{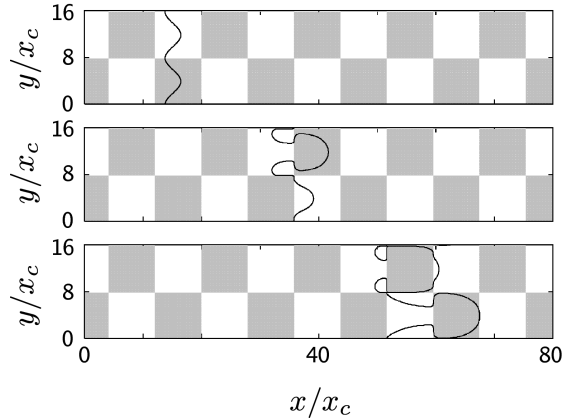}\\
\end{centering}
\caption{Contact line growth on a symmetric checkerboard pattern.  The length and width of the domains is fixed to 
$\Lambda_{max}(\theta_e=90^\circ)$.  The first panel shows the initial perturbation to the front at $t/t_c=0$ with a perturbation 
wavelength that equals $\Lambda_{max}(\theta_e=90^\circ)$.  The next panels show the contact line 
at time intervals of $\Delta t /t_c= 26.8$.  For both upper and lower domains, the contact line grows as a finger when in contact with a hydrophobic domain, 
and spreads when in contact with a hydrophilic domain.  Overall, the net growth rate of the contact line achieves a constant value.\label{fig:cbpattern2}}
\end{figure}

Controlled tuning of the growth rate of the film is an appealing technique, that could be used, for instance,  to control 
the filamenting of the film in chemical networks, where different fluid filaments move at different velocities
depending on the growth rate on a specific chemical stripe.  The results presented in the previous section indicate that 
a thin film can be oriented along an imposed path by using longitudinal striped patterns,  thus favoring filamenting.
In addition, such patterns are useful to control the spacing between emerging filaments. 
In this section we propose a way to control the growth rate of the film by studying the effect 
of a pattern consisting of hydrophilic and hydrophobic domains arranged in an asymmetric checkerboard configuration, 
as displayed in  figure~\ref{fig:cbpattern}.   Such a pattern converges to the longitudinal pattern the limit where 
the length of the hydrophilic or hydrophobic domains, $l_h$ and $l_\phi$, is large enough, thus giving us the possibility
of direct comparison to results presented in the last section. 

The checkerboard  pattern introduces a transverse spreading mechanism, arising
every time that the film comes into contact with a hydrophilic domain, and that appears as a possible means to control 
the motion of the contact line.  As before, we are interested in the interplay between the lengthscales of the chemical 
pattern and the intrinsic lengthscale of the contact line.  We will therefore consider situations in which the width 
of the lower domains shown in the figure is fixed to $w_\phi \simeq \Lambda_{max}(\theta_e=90^\circ)$, while the width of the 
upper domains is varied.  In this way, it is expected that the contact line develops as a finger on the lower domains, if the domain is 
hydrophobic, and spreads sideways to form a sawtooth if the domain is hydrophilic.   Meanwhile, if the width of the upper domains is 
sufficiently small, the contact line is not expected to deform appreciably. However, for wide enough upper domains, a fingering-spreading 
process of the film, as on the lower domains, is expected.
 
Choosing the width of the thinner domains, $w_h$, as well as the  lengths of  the hydrophilic and hydrophobic domains, $l_h$ and $l_\phi$, 
fixes the fraction of the substrate composed of hydrophilic material, $f_{CB}=(w_hl_\phi+w_\phi l_h)/[(l_h+l_\phi)(w_h+w_\phi)]$.  
Given that the widths $w_h$ and $w_\phi$ are fixed, the limiting cases of $l_\phi=0$ and $l_\phi \rightarrow \infty$ correspond to  
patterns of adjacent longitudinal stripes, as the ones shown in figures~\ref{fig:lstripes_small}(a) and \ref{fig:lstripes_small}(b).   In the previous 
section we found that contact line growth is observed in these patterns.  Consequently, at intermediate values of $l_\phi$, growth 
is expected for checkerboard patterns.

We first consider the case in which the front can destabilize only on the wide domains, so we fix $w_h$  to a value that is much smaller than the most 
unstable wavelength.  To gain insight on the effect of the checkerboard pattern on the dynamics of the contact line, we
consider the effect of varying the fraction of hydrophilic domains on the evolution of the thin film.  We do this by fixing the 
length of the hydrophilic domains to an arbitrary value, which we choose as $l_h/x_c=1.3$, while the length of hydrophobic 
domains is varied.  The width of the upper and lower domains is fixed to $w_h/x_c = 1.3$ and $w_\phi=8$, respectively.  
We consider six values of  $l_\phi$, namely,  $l_\phi/x_c=0.6$, $l_\phi/x_c=1.0$, $l_\phi/x_c=1.6$, $l_\phi/x_c=2.7$, $l_\phi/x_c=4.0$ 
and $l_\phi/x_c=8.0$. The front is perturbed initially on the wider domains using the most unstable 
wavelength, and on the thinner domains using a wavelength equal to $w_h$.  We follow the evolution 
of the total length of the contact line, $L$, by measuring its
growth rate, $\dot L=\dot L (w_\phi,w_h,l_\phi,l_h)$, which is calculated 
as the difference between the leading and trailing edge velocities.  The leading edge  
is located at the middle of the wider domains, while the trailing edge is taken as 
the contact line position at the middle of the 
thinner domains, as shown in figure \ref{fig:cbpattern}.

The evolution of the front is tracked in figures~\ref{fig:cbpattern1}(a) and~\ref{fig:cbpattern1}(b)  for $l_\phi/x_c=0.6$, and $l_\phi/x_c=8$,  which correspond to hydrophilic fractions $f_{CB}=0.25$ and $f_{CB}=0.62$, respectively.     
In these figures we also show the finger that grows on  a homogeneous substrate.  For both checkerboard patterns, the contact line grows as a finger on a hydrophobic domain, then spreads out on a hydrophilic domain, and finally 
grows again as a finger as it touches the next hydrophobic domain.  At a given time, comparing the contact line 
for both $l_\phi$ values shows that the leading edge is located at a slightly more advanced position for the smaller $l_\phi$, whereas the trailing edge is located at a slightly less advanced position.  This means that the leading edge advances faster for small $l_\phi$.  This occurs because the resistance to contact line motion is smaller in this case, given that $f_{CB}$ is large. Conversely, at small $l_\phi$, the trailing edge has to sweep increasingly long hydrophobic domains, thus decreasing its velocity.  Still, whatever the value of $l_\phi$, the comparison to the homogeneous substrate shown in figure~\ref{fig:cbpattern1} 
clearly shows that the contact line grows to smaller lengths as a consequence of the checkerboard pattern. 
In figure~\ref{fig:ldotcheckerboard} we plot the growth rate of the contact line as a function of the length of the hydrophobic domains.  The limiting cases $\dot L (l_\phi=0)$ and $\dot L (l_\phi \rightarrow \infty)$ correspond to the growth rates of the contact line in the adjacent stripe patterns of figures~\ref{fig:lstripes_small}(b) 
and~\ref{fig:lstripes_small}(a), given that $w_h$ and $w_\phi$ are chosen to be equal to the stripe widths of those patterns.  
For intermediate $l_\phi$, we observe a monotonous decrease from one value to the other.

Increasing the width of the thin domains, $w_h$, does not modify the dynamics of the film qualitatively, 
as long as this width is much smaller than $\Lambda_{max}$.    This changes as $w_h$
becomes comparable to $\Lambda_{max}$; in this case the contact line destabilizes in both domains, giving rise to the growth of a different structure.
Figure~\ref{fig:cbpattern2} shows the contact line growth on equally wide domains, of width equal to $\Lambda_{max}$.  In this case, there are two fronts that 
grow with the same velocity.  Each front grows as a finger when in contact with a hydrophobic domain,  and spreads on the following hydrophilic domain 
periodically.  At long times, the net growth rate of  the contact line achieves a constant value.   

Thus, for both symmetric and asymmetric configurations, the checkerboard pattern introduces the effect of spreading 
on hydrophilic domains, relaxing the growth of a finger momentarily.  For asymmetric patterns, this provides 
a way to tune the growth rate of the contact line.  In the following section we will propose a kinematical model 
to account for the growth of the contact line in terms of the geometrical parameters of the checkerboard pattern.  

\subsubsection*{Kinematical model for growth on checkerboard patterns}
\begin{figure}
\begin{centering}
\includegraphics[width=0.45\textwidth]{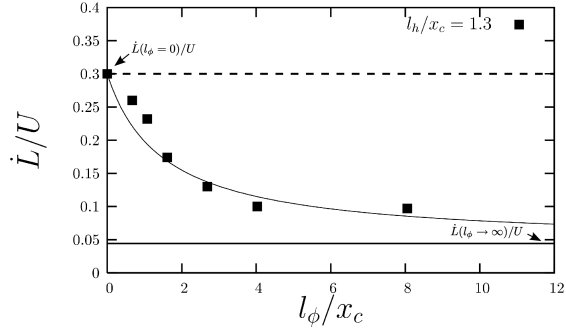}\\
\end{centering}
\caption{Contact line growth rate as a function of hydrophobic domain length on checkerboard patterns.  The growth rate decreases 
as the length of the hydrophobic domains increases.  This effect is caused by the adjacent thin hydrophilic domains, whose length 
increases with $l_\phi$, thus making the trailing edge velocity larger and $\dot L $ smaller.   The solid curve corresponds to the theoretical
prediction,  equation~(\ref{eq:dotLCB}).  The straight lines correspond to the limiting growth rates
$\dot L (l_\phi=0)$ (dashed line) and $\dot L (l_\phi\rightarrow \infty)$ (solid line).  \label{fig:ldotcheckerboard}}
\end{figure}

To quantify the effect of the checkerboard pattern on the contact line growth, we characterize 
the motion of the thin film in terms of the velocities of the leading an trailing edges 
in the wide and thin domains.  In general, the instantaneous leading and trailing edge 
velocities, $v_l$ and $v_t$,  will depend on the configuration of the pattern, {\it i.e.}, $v_l=v_l(w_\phi,w_h,l_h,l_\phi)$ and $v_t=v_t(w_\phi,w_h,l_h,l_\phi)$.   We assume 
that these velocities relax to constant values as soon as the contact line comes into contact with a hydrophilic or a hydrophobic domain.   For
 simplicity, we take these values as the ones corresponding to the limiting cases of very long domains. 
Therefore, at a hydrophobic domain,  the leading and trailing edge velocities are approximated as
 $v_l^\phi=v_l(w_\phi,w_h,l_h=0,l_\phi\rightarrow\infty)$ and
 $v_t^\phi=v_t(w_\phi,w_h,l_h\rightarrow\infty,l_\phi=0)$, while at a hydrophilic domain we have
 $v_l^h=v_l(w_\phi,w_h,l_h \rightarrow \infty,l_\phi=0)$ and  $v_t^h=v_t(w_\phi,w_h,l_h=0,l_\phi\rightarrow\infty)$ .  

We denote $\Delta t_l$ and $\Delta t_t$ the times in which the leading and trailing edges 
sweep a period of the underlying pattern, $l_h+l_\phi$.  It follows that
\begin{equation}
\Delta t_l = \Delta t_l^h + \Delta t_l^\phi \qquad  \mathrm{and} \qquad \Delta t_t = \Delta t_t^h + \Delta t_t^\phi. 
\label{eq:CBtimes}
\end{equation}
In these expressions, the subscripts $l$ and $t$, stand for the leading and trailing
edges, while the superscripts $h$ and $\phi$ stand for hydrophilic and hydrophobic 
domains respectively.   The leading edge will sweep a hydrophilic domain of length $l_h$
with a velocity $v_l^h$ and will then move across a hydrophobic domain of length $l_\phi$
with a velocity $v_l^\phi.$  On the other hand, because of the geometry of the checkerboard pattern, 
the trailing edge will move over a hydrophilic domain of length $l_\phi$ with a velocity $v_t^h$, to later continue over a hydrophobic domain of length $l_h$ with a velocity $v_t^\phi$. 

Accordingly, equation~(\ref{eq:CBtimes}) can be written as
\begin{equation}
\frac{l_\phi+l_h}{\hat v_l}= \frac{l_h}{{v_l}^h}+\frac{l_\phi}{{v_l}^\phi}\qquad \mathrm{and} \qquad \frac{l_\phi+l_h}{\hat v_t} = \frac{l_\phi}{{v_t}^h}+\frac{l_h}{{v_t}^\phi},
\label{eq:CBvels}
\end{equation}
in terms of the mean values of the leading an trailing edges, $\hat v_l$ and $\hat v_t$, respectively.  We can finally compute the 
average growth rate of the contact line as $\dot L = \hat v_l - \hat v_t,$ which within approximations considered reads
\begin{equation}
\dot L = \left(l_\phi + l_h\right)\left[\left(\frac{l_h}{{v_l}^h}+\frac{l_\phi}{{v_l}^\phi}\right)^{-1}-\left(\frac{l_h}{{v_t}^\phi}+\frac{l_\phi}{{v_t}^h}\right)^{-1}\right].
\label{eq:dotLCB}
\end{equation}
Equation~(\ref{eq:dotLCB}) gives an estimate of the growth rate of the contact line in terms of the velocities observed in longitudinal striped patterns, the width of the stripes being equal to the width of the domains of the checkerboard 
arrangement.     To compare simulation results with the kinematical prediction, we measure 
$v_l^\phi$, $v_l^h$, $v_t^\phi$, and $v_t^h$ from runs that correspond to a longitudinal stripe configuration 
for which $w_h$ and $w_\phi$ match with the checkerboard pattern, as displayed in
 figures~\ref{fig:lstripes_small}(a) and~\ref{fig:lstripes_small}(b).  Figure~\ref{fig:ldotcheckerboard} 
shows a comparison between the growth rate measured from simulations and the kinematical prediction
 as a function of  the length of the hydrophobic domains. The quantitative agreement shows
 that the varying wettability of alternating patches determines the growth of the contact line essentially
by fixing the local value of the leading and trailing edge velocities, these values being very well 
approximated by the ones corresponding to long domains. 

\section{Discussion and Conclusions}
\label{sec:DC}
By means of lattice-Boltzmann simulations and kinematical models,  
we have studied the dynamics of driven thin films on a variety of heterogeneous 
hydrophilic-hydrophobic  substrates.   We have studied the effect of longitudinally
stripes and checkerboard patterns.  

We have focused on a scenario where the unstable contact line gives rise to 
sawtooth and finger structures on homogeneous hydrophilic and hydrophobic substrates,
respectively, and have examined how the growth of the contact line  is altered by the 
chemical pattern.    To this end, we have considered patterns where the typical lengthscale is comparable
to the most unstable wavelength of the contact line, $\Lambda_{max}$.  

On longitudinal patterns, the film follows hydrophilic stripes preferentially, while 
the contact line gives rise to fingering if the width of the hydrophobic stripes is 
large enough.  For small enough hydrophobic stripes, the film gives rise to a fingering
process caused by the longitudinal pattern, where the width of the growing fingers
corresponds to the width of the hydrophilic stripes.  

On checkerboard patterns, the film undergoes a fingering-spreading process as 
it moves over hydrophobic and hydrophilic domains.  Using a kinematical approach, 
we have shown that the net growth rate of the contact line can be tuned by choosing
the lengthscale of the checkerboard pattern. 

In conclusion, we have performed lattice-Boltzmann simulations of the evolution
of a three dimensional thin film in contact with chemically patterned substrates
We have demonstrated that the film can be oriented along a given path, that fingering 
can be triggered by the chemical pattern, and that the growth of the contact line can be controlled 
by choosing a particular configuration for substrate patterning.  

\label{sec:Conclusions}

\section{Acknowledgments}
R.L-A. wishes to thank M. Pradas for useful discussions.  We acknowledge financial support from Direcci\'on General de Investigaci\'on (Spain) under projects FIS\ 2009-12964-C05-02 and FIS\ 2008-04386.   R.L.-A. acknowledges support from CONACyT (M\'exico) and Fundaci\'on Carolina(Spain).    The computational work presented herein has been carried out in the MareNostrum Supercomputer at Barcelona Supercomputing Center.  \label{sec:Acknowl}

\section{Supporting Information Available}
This information is available free of charge via the Internet at http://pubs.acs.org/.

\providecommand*{\mcitethebibliography}{\thebibliography}
\csname @ifundefined\endcsname{endmcitethebibliography}
{\let\endmcitethebibliography\endthebibliography}{}

\end{document}